\documentclass[runningheads]{llncs}
\usepackage[T1]{fontenc}
\usepackage{graphicx,verbatim}
\usepackage{amsmath}
\usepackage{amssymb}
\usepackage{booktabs} 
\usepackage[table]{xcolor}
\usepackage{multirow}
\begin{document}
\title{A Benchmark of (MRI-) Foundation Models to Predict IDH Mutational Status in Glioma}
%\titlerunning{Abbreviated paper title}
% If the paper title is too long for the running head, you can set
% an abbreviated paper title here
%
\author{Nathan Hollet\inst{1,2,*} \and
Elise Robinson\inst{1,3,*} \and
Efthymios Georgiou\inst{1, 2} \and
Ekin Ermis\inst{1} \and
Uri Nahum\inst{4,5} \and
Sarah Brüningk\inst{1,2} }
\authorrunning{N. Hollet et al.}
% First names are abbreviated in the running head.
% If there are more than two authors, 'et al.' is used.
%
\institute{Department of Radiation Oncology, Inselspital, Bern University Hospital and University of Bern, Switzerland \and Department of Digital Medicine, University of Bern, Switzerland \and School of Life Sciences, FHNW University of Applied Sciences and Arts, Muttenz, Switzerland \and School of Business, FHNW University of Applied Sciences and Arts, Olten, Switzerland \and University Children's Hospital Basel UKBB, Basel, Switzerland\\
\email{\{nathan.hollet, efthymios.georgiou, sarah.brueningk\}@unibe.ch}\\
\email{elise.robinson@students.fhnw.ch}\\
\email{ekin.ermis@insel.ch} \\
\email{uri.nahum@fhnw.ch}}
  
\maketitle              % typeset the header of the contribution
\let\oldthefootnote\thefootnote
\renewcommand{\thefootnote}{}%
\footnotetext{*: shared first author}
\renewcommand{\thefootnote}{\oldthefootnote}%

\begin{abstract}
Non-invasive prediction of glioma molecular status from routine magnetic resonance imaging (MRI) has shown promising performance, but model generalization remains challenging given small-scale matched imaging-genomic datasets. Foundation models may address this bottleneck, but a comprehensive benchmark is needed to establish the impact of diverse architectures, pre-training domains, and objectives. Given the use case of isocitrate dehydrogenase (IDH) mutation prediction from FLAIR and post-contrast T1 MRIs, we compared four image-based foundation models, BrainIAC, MRI-CORE, BiomedCLIP, and BrainDINO, against radiomics-based TabPFN and logistic regression baselines. Prediction performance and calibration were assessed across four public adult glioma cohorts and an external post-treatment cohort. Within-cohort, TabPFN matched or outperformed all visual encoders, achieving 0.92 (0.03) AUROC and 0.74 (0.17) AUPRC (mean (SD) across all datasets). Among visual encoders, BiomedCLIP performed best (0.85 (0.08) AUROC), with BrainDINO competitive (0.82 (0.09) AUROC), while MRI-specific encoders (BrainIAC, MRI-CORE) consistently underperformed. Cross-cohort transfer showed moderate AUROC degradation but stronger AUPRC sensitivity to prevalence shifts. On the external cohort, BiomedCLIP achieved the highest AUROC (0.74 (0.07)), whereas TabPFN provided superior calibration (Expected Calibration Error 0.07 (0.01)). These results indicate that representation modality and evaluation context critically influence foundation-model performance in MRI-based molecular prediction. Tabular foundation models on radiomic features provide a strong, well-calibrated baseline, while image foundation models may offer complementary value under clinically distinct distribution shifts. The code will be made available upon acceptance.


% Authors must provide keywords and are not allowed to remove this Keyword section.

\end{abstract}

\section{Introduction}
Effective molecular characterization of brain tumors is essential for treatment selection and pre-operative planning. The clinical standard for molecular profiling relies on invasive biopsies \cite{ziv_importance_2016}, which are costly and prone to sampling errors \cite{sottoriva_intratumor_2013}.  Magnetic resonance imaging (MRI) is routinely acquired \cite{van_der_heide_mri_2019}, making MRIs a readily available, non-invasive data source to infer molecular characteristics. The isocitrate dehydrogenase (IDH) mutation status is of particular importance, as it serves as a key classifier in the current World Health Organization (WHO) grading system, carrying strong prognostic implications and guiding treatment decisions \cite{yan_idh1_2009,louis_2021_2021}. Predicting IDH status from MRI would therefore have direct clinical value. Recent studies have extensively explored the use of molecular prediction methods, including radiomics and supervised deep learning. Handcrafted radiomics-based approaches extract quantitative features such as texture, shape, and intensity statistics from tumor ROIs on MRI scans and feed them into machine learning classifiers \cite{lu_noninvasive_2023}, achieving competitive performance but limited generalization \cite{calabrese_combining_2022,choi_fully_2020,an_radiomics_2021}. Deep learning methods, particularly CNNs, have further improved performance but remain limited by the small sample size of typical matched MRI and genomic datasets \cite{he_radiogenomics_2024}. Consequently, deep feature extractors were outperformed by radiomics on moderate-sized cohorts \cite{nakase_integration_2025}. Moreover, a recent systematic review of deep-learning IDH prediction \cite{farahani_diagnostic_2026} highlights extensive multi-center external validation as a critical missing step toward clinical translation. This motivates the use of vision foundation models (VFMs), which leverage large-scale self-supervised pretraining to learn representations transferable via lightweight probing even when labeled data is scarce. \cite{bommasani_opportunities_2022}. VFMs capture broad visual features transferable to downstream tasks via fine-tuning or lightweight probing, even when labeled data is scarce \cite{oquab2024dinov2learningrobustvisual}. In the medical imaging domain, several VFMs have recently been proposed \cite{zhang_biomedclip_2025,dong_mri-core_2025,tak_generalizable_2026,wu2026braindinobrainmrifoundation}, each trained on different types and scales of data, ranging from modality-specific (Brain-specific or MRI-specific) to more general collections of biomedical images. Importantly, the trade-off between broad pretraining data and smaller-scale task-specific training has not yet been systematically evaluated. This study aims to investigate the impact of pre-training domain, feature representation, and evaluation setting on molecular prediction performance and generalization. We present a systematic comparison of four image-based foundation models spanning different pre-training domains: BrainIAC \cite{tak_generalizable_2026}, pre-trained on brain MRI; MRI-CORE \cite{dong_mri-core_2025}, pre-trained on multi-anatomy MRI slices; BiomedCLIP \cite{zhang_biomedclip_2025}, pre-trained contrastively on biomedical image--text pairs; and BrainDINO \cite{wu2026braindinobrainmrifoundation} pre-trained on 6.6M brain MRI slices. Using IDH mutation status prediction from FLAIR and post-contrast T1-weighted MRI as a representative prediction task, we evaluate these models under within-cohort, cross-cohort, and external post-treatment evaluation. We contextualize against radiomics-based baselines, including a tabular foundation model applied to hand-crafted imaging features.

\section{Methods}
\begin{figure}[ht!]
    \centering
    \includegraphics[width=\textwidth]{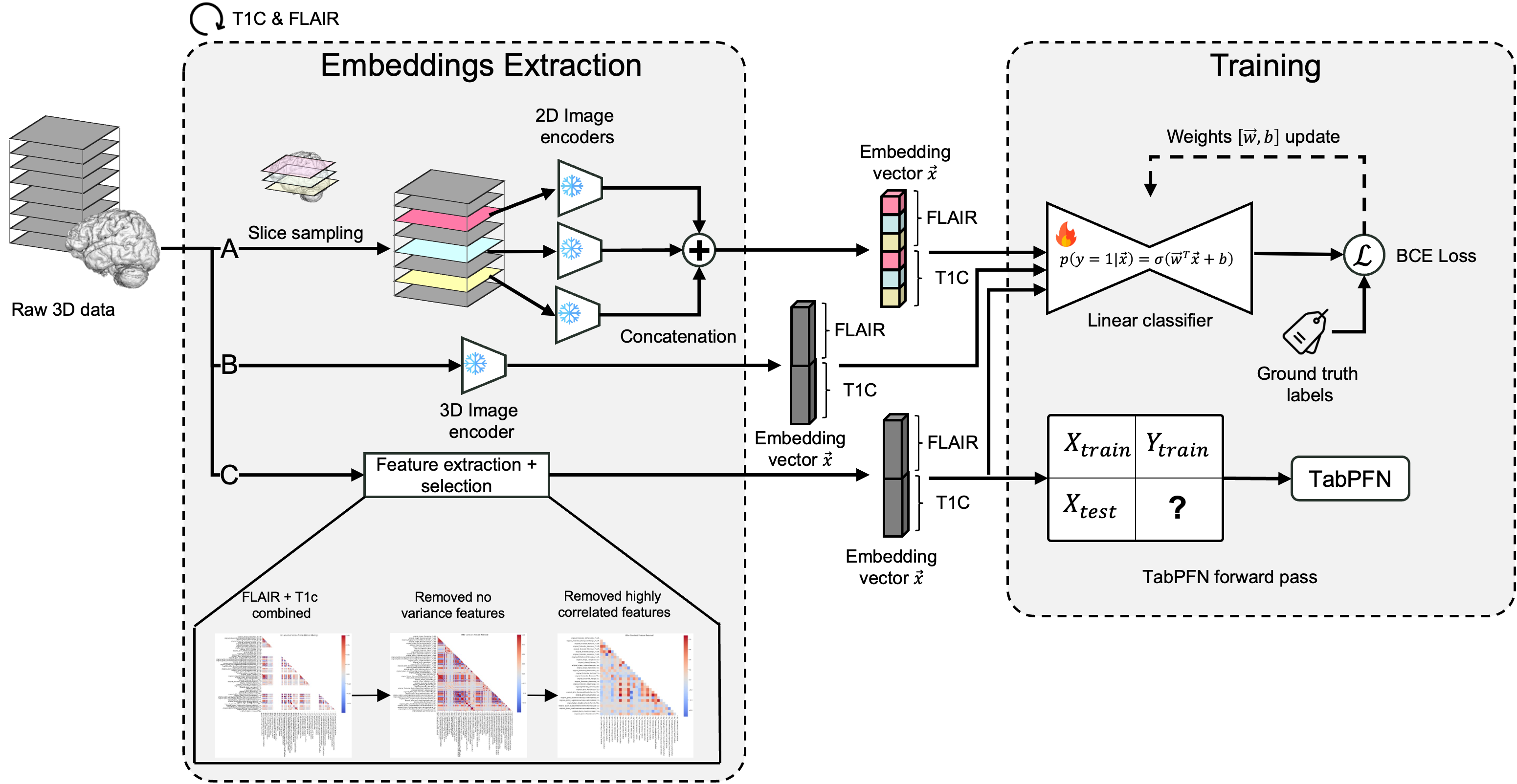}
    \caption{Schematic overview. Each scan (FLAIR and post-contrast T1) feeds three parallel feature extraction pipelines. (A) 2D image foundation models (MRI-CORE, BiomedCLIP, BrainDINO)Three axial slices at the 25th/50th/75th percentiles of the tumor mask pass independently through a frozen encoder, with CLS tokens concatenated. (B) 3D image foundation model (BrainIAC): the full volume passes through a frozen encoder. (C) Radiomics: PyRadiomics descriptors with per-fold feature selection (constant + highly correlated features dropped). All vectors feed a trainable linear classifier under BCE loss (equivalent to logistic regression for the radiomics path); the radiomics vector additionally feeds TabPFN in-context with no gradient updates. Snowflakes denote frozen modules, flames trainable.}
    \label{fig:pipeline}
\end{figure}

We separate the training workflows into two independent parts: baseline and linear probe training, as summarized in Figure~\ref{fig:pipeline}. Both workflows operate on identical patient cohorts and share the same train/validation/test splits and cross-cohort evaluation protocol.

\subsection{Datasets} \label{datasets}
Cohort characteristics are summarized in Table~\ref{tab:cohorts}. All cohorts arrive skull-stripped and co-registered at 1 mm isotropic resolution, except EGD, which was not skull-stripped (processed per-sequence with HD-BET~\cite{isensee_automated_2019}). UCSD-PTGBM is post-treatment by design: 85\% of the 123 patients had prior surgery (52\% gross total, 33\% subtotal, 15\% biopsy-only), providing a clinically realistic out-of-distribution test of preoperatively trained models against post-resection cavities and treatment-induced changes. T2-FLAIR and T1c are used throughout; tumor segmentations and IDH labels are taken as provided.

\begin{table}[ht!]
  \centering \small
  \caption{Cohort summary. All cohorts use T2-FLAIR + post-contrast T1, arrive skull-stripped and co-registered at 1mm isotropic (EGD: not skull-stripped, processed with HD-BET~\cite{isensee_automated_2019}). $^\dagger$EGD: +52 grade-unknown.}
  \label{tab:cohorts}
  \setlength{\tabcolsep}{3pt}
  \begin{tabular}{lrrrrrr}
    \toprule
    \bfseries Cohort & \bfseries N & \bfseries Years & \bfseries Grade 2 & \bfseries Grade 3 & \bfseries Grade 4 & \bfseries \% IDH-mut \\
    \midrule
    UCSF-PDGM~\cite{calabrese_university_2022}   & 494 & 2015--21 & 56  & 43 & 395 & 21\%  \\
    UPENN-GBM~\cite{bakas_university_2022}       & 515 & 2006--18 & 0   & 0  & 515 & 3.1\% \\
    EGD~\cite{van_der_voort_erasmus_2021}        & 467 & 2008--19 & 127 & 29 & 259$^\dagger$ & 33\%  \\
    UTSW-Glioma~\cite{reddy_university_2026}     & 619 & --       & 100 & 97 & 416 & 28\%  \\
    UCSD-PTGBM~\cite{gagnon_university_2025}     & 123 & --       & 0   & 0  & 123 & 9.8\% \\
    \bottomrule
  \end{tabular}
\end{table}

\subsection{Baselines} \label{baselines}
\noindent \textbf{Radiomics feature extraction}
Both radiomics baselines (logistic regression and TabPFN, Table~\ref{tab:models}) share the same input: a per-patient feature vector formed by concatenating PyRadiomics (v3.1.1) descriptors extracted from the segmented tumor on FLAIR and post-contrast T1. The extractor's default configuration is used, yielding 107 PyRadiomics features per image (14 shape, 18 first-order, and 75 texture features across GLCM, GLRLM, GLSZM, GLDM and NGTDM), concatenated across FLAIR and post-contrast T1 for a 214-d per-patient vector before feature selection. The vector is pruned per fold exclusively on the training split by dropping constant columns and one of each pair of strongly correlated features (Pearson $|\rho| > 0.9$), discarding the one with the weaker absolute correlation to the training label.

\noindent \textbf{Logistic regression}
The first classifier is an $\ell_1/\ell_2$-regularized logistic regression \cite{qin_l1-2_2019}, fitted by minimizing a class-weighted cross-entropy (class weights inversely proportional to class frequency, compensating for label imbalance) with an additive $\ell_1$ or $\ell_2$ penalty controlled by the regularization strength $C$. 

\noindent \textbf{TabPFN}
The second is TabPFN \cite{hollmann2023tabpfn}, a transformer pretrained once on a synthetic prior over tabular classification tasks and used without further training: rather than performing gradient updates, it approximates the posterior predictive by in-context learning, conditioning a single frozen forward pass on the labelled training rows to output the IDH probability for each query patient.

\subsection{Image-based foundation models} \label{foundation}
Evaluating the image encoders (Table~\ref{tab:models}) as frozen feature extractors with a linear probe within an identical pipeline ensures that performance differences reflect the pre-trained representations rather than the training procedure. As summarized in Table~\ref{tab:models}, the four image encoders span the design space of brain-specific vs. broader pretraining and of contrastive vs. self-distillation objectives.

\begin{table}[ht!]
  \centering
  \small
  \setlength{\tabcolsep}{4pt}
  \renewcommand{\arraystretch}{1.2}
    \caption{Foundation models benchmarked. All image encoders are ViT-B/16 variants (BrainIAC: 3D; MRI-CORE: SAM-style); TabPFN: tabular transformer. \emph{Input vec.}: per-patient FLAIR + T1 vector dimension fed to the linear probe.}
  \label{tab:models}
  \begin{tabular}{llrrr}
    \toprule
    \bfseries Model & \bfseries Data & \bfseries Obj. & \bfseries Params & \bfseries  Input vec.\\
    \midrule
    BrainIAC \cite{tak_generalizable_2026}       & 32k brain MRI                      & SimCLR \cite{chen2020simpleframeworkcontrastivelearning} & 88.3 M     & 1{,}536 \\
    MRI-CORE \cite{dong_mri-core_2025}           & 110k MRI, 18 anatomies             & DINOv2 \cite{oquab2024dinov2learningrobustvisual}        & 85.8 M     & 1{,}536 \\
    BiomedCLIP \cite{zhang_biomedclip_2025}      & 15M image--text pairs              & CLIP \cite{radford_learning_2021}                        & 86.2 M     & 3{,}072 \\
    BrainDINO \cite{wu2026braindinobrainmrifoundation} & 6.6M brain MRI slices & DINOv3 \cite{simeoni2025dinov3}                          & 85.6 M     & 4{,}608 \\
    TabPFN \cite{hollmann2023tabpfn}             & synthetic tabular tasks            & PFN \cite{hollmann2023tabpfn}                            & $\sim$28 M & 214 \\
    \bottomrule
  \end{tabular}
\end{table}

\noindent \textbf{Image embedding extraction}
Embeddings are extracted once per patient and modality (FLAIR and post-contrast T1), then cached. BrainIAC is 3D and consumes the whole skull-stripped volume, returning one feature vector per modality. MRI-CORE, BiomedCLIP, and BrainDINO are 2D, so we extract embeddings from three axial slices at the $25$th, $50$th, and $75$th percentile of the tumour mask along the cranio-caudal axis and concatenate them. The FLAIR and T1 vectors are then concatenated to produce the input to the probe.

\noindent \textbf{Linear probe training}
The probe is a single linear layer trained by minimizing a weighted binary cross-entropy, with the positive class up-weighted by $\alpha = \min(N_{-}/N_{+}, 10)$ and the inverse class ratio capped at 10 to prevent gradient blow-up on cohorts with very few positives. The linear probe (PyTorch \cite{paszke2019pytorchimperativestylehighperformance}) is trained with Adam (lr $\in {10^{-5},10^{-4},10^{-3}}$, weight decay $10^{-4}$, batch 32, 500 epochs, early stopping after 30 epochs on AUPRC). Logistic regression is grid-searched over $C \in {10^{-4},\dots,10^{2}}$ and penalty $\in {\ell_1,\ell_2}$ (LIBLINEAR, $\leq 2000$ iters). TabPFN uses defaults. Hyperparameters are selected on validation AUPRC.

\subsection{Evaluation protocol} \label{evaluation}
For each cohort, we generate $K = 5$ independent train/val/test splits at $70/15/15$ by multi-label stratified shuffle sampling on (label, WHO grade, age tertile), so each run yields an independently stratified triple and paired comparisons operate on identical patient lists (test sets may overlap across runs). Within-cohort performance is summarized as mean $\pm$ std over the $K$ runs for $m \in \{\mathrm{AUROC}, \mathrm{AUPRC}\}$, both computed directly from predicted probabilities. For each ordered pair $(A, B)$ with $A \neq B$, the model trained on $A$ in run $i$ is applied to the entirety of $B$ without refitting. Radiomics baselines replay the per-run feature selection fitted on $A$ on $B$, and foundation-model baselines use embeddings pre-extracted by the same frozen encoder. The diagonal of the cross-cohort heatmap is therefore computed on the held-out test split of cohort $A$, while off-diagonal entries are computed on the full cohort $B$. To assess probability calibration, we additionally compute the Expected Calibration Error (ECE) with 10 quantile bins: for each bin, the absolute difference between mean predicted confidence and observed accuracy, weighted by bin size.

\section{Results}

\begin{figure}[ht!]
    \centering
    \includegraphics[width=\textwidth]{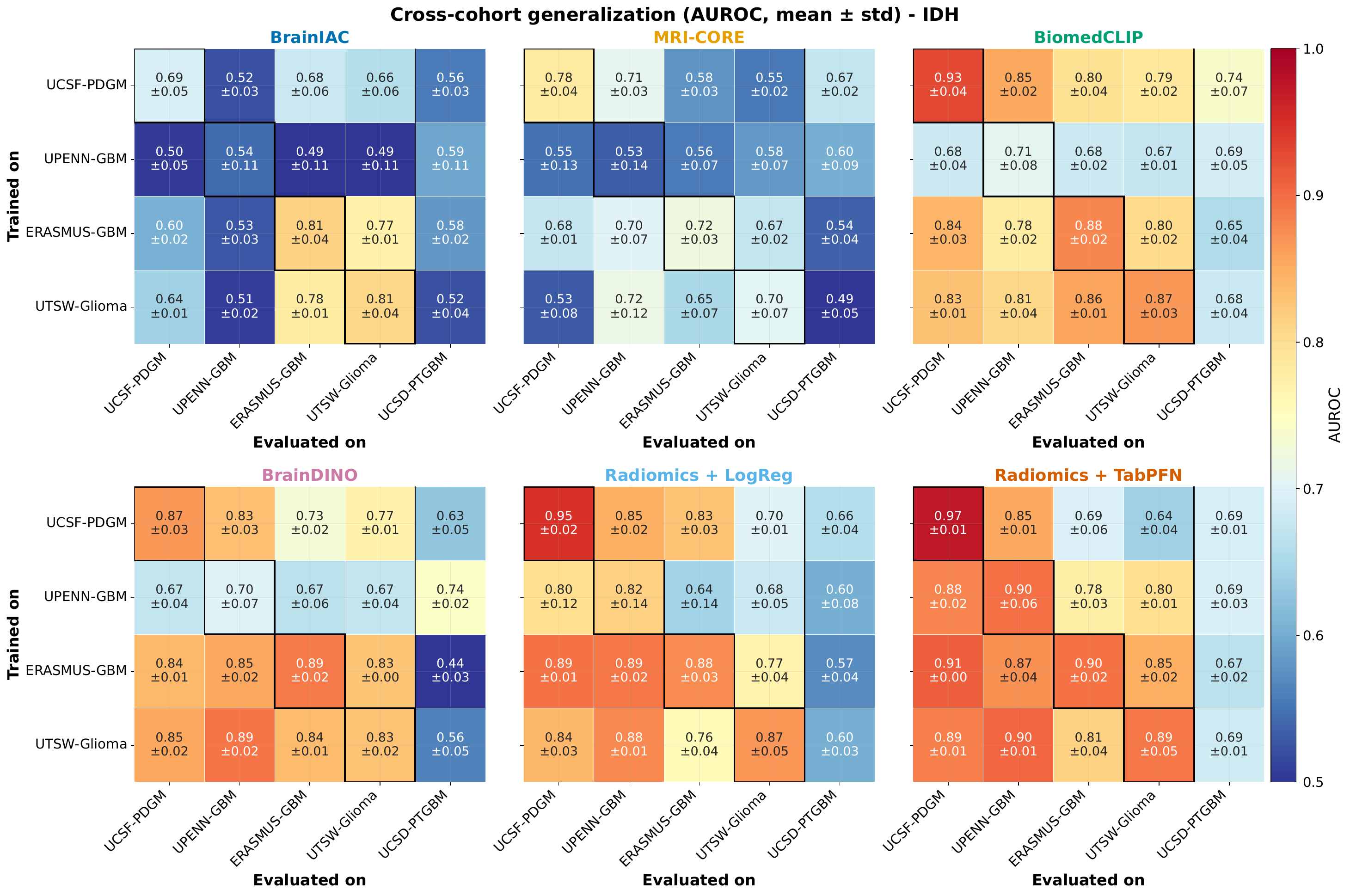}
    \caption{Cross-cohort AUROC (mean $\pm$ std). Rows are training cohorts, columns evaluation cohorts; the boxed diagonal cells are the within-cohort numbers.}
    \label{fig:gen_heatmap_auroc}
\end{figure}

We report performance across the six approaches using $K = 5$ paired runs per cohort, as mean $\pm$ std throughout. We report AUROC throughout; AUPRC was evaluated identically and follows the same model ordering except where noted.

\noindent\textbf{Within-cohort performance.}
The boxed diagonal of Figure~\ref{fig:gen_heatmap_auroc} summarizes in-distribution performance. Averaged across the four training cohorts, TabPFN ranks first (AUROC $0.92 \pm 0.03$, AUPRC $0.74 \pm 0.17$), followed by LogReg ($0.88 \pm 0.05$ / $0.68 \pm 0.22$). BiomedCLIP is the strongest imaging foundation model on three of four training cohorts (UCSF, UPENN, UTSW), with BrainDINO close behind in second; on EGD, BrainDINO narrowly overtakes BiomedCLIP ($0.889$ vs $0.880$) and is the top imaging encoder. The current MRI-specific encoders (MRI-CORE, BrainIAC) consistently rank last or near-last. AUPRC follows the same ordering on UCSF, UPENN, and UTSW, but on EGD TabPFN ($0.783$) is overtaken by both BiomedCLIP ($0.784$) and BrainDINO ($0.822$).\\
\noindent\textbf{Cross-cohort generalization.}
The off-diagonal cells of Figure~\ref{fig:gen_heatmap_auroc} show that AUROC transfer between the four training cohorts is broadly preserved: most drops fall in the $0.05$--$0.15$ range for the top baselines, and a handful of cells match or slightly exceed the source diagonal (e.g.\ EGD~$\to$~UCSF-PDGM for the radiomics baselines). AUPRC drops are substantially larger, especially in the low-prevalence UPENN-GBM ($3.1\%$ mutant), reflecting AUPRC's sensitivity to class composition. UPENN-GBM is also the weakest training cohort: with only $16$ mutant cases, its variance is high, and models trained on it transfer worst on AUPRC. Averaged over all cross-cohort transfers, TabPFN remains the top-ranked baseline; BiomedCLIP overtakes it only on the most demanding shift. \\
\noindent\textbf{External cohort (UCSD-PTGBM).}
On the external UCSD-PTGBM cohort, every baseline drops $0.10$--$0.30$ AUROC and up to $0.69$ AUPRC relative to its source diagonal. The in-distribution ranking reverses on this external test: BiomedCLIP and BrainDINO achieve the highest external AUROC ($0.740$ for BiomedCLIP trained on UCSF-PDGM; $0.743$ for BrainDINO trained on UPENN-GBM), while the radiomics-based models drop more sharply. \\
\noindent\textbf{Calibration.}
To complement these ranking-based metrics, we also assess probability calibration: a model can rank correctly while producing untrustworthy probability outputs. Table~\ref{tab:calibration_ece} reports the Expected Calibration Error (ECE; lower is better). TabPFN is the best-calibrated baseline in every setting (in-distribution ECE $0.03$--$0.10$, $0.094$ averaged across cross-cohort transfers, and $0.073$ on the external UCSD-PTGBM cohort). BiomedCLIP and BrainDINO are consistently next in-distribution (ECE $0.09$--$0.22$), with BrainDINO actually edging out BiomedCLIP on EGD ($0.10$ vs $0.11$) and UTSW ($0.12$ vs $0.13$); the current MRI-specific encoders (BrainIAC, MRI-CORE) and LogReg are moderately to poorly calibrated (ECE $0.15$--$0.32$). On the external UCSD-PTGBM cohort, however, BrainDINO's calibration degrades sharply to $0.26$, falling behind BiomedCLIP ($0.13$) and even LogReg ($0.18$), mirroring its AUROC pattern of strong within-domain performance but weaker robustness under heavy distribution shift. Notably, the BiomedCLIP AUROC advantage on UCSD-PTGBM does not translate into a calibration advantage: TabPFN ranks second on AUROC there but remains the best-calibrated baseline.

\begin{table}[ht!]
    \centering
    \small
    \setlength{\tabcolsep}{3pt}
    \renewcommand{\arraystretch}{1.1}
    \caption{Expected Calibration Error (ECE $\downarrow$; lower is better; mean $\pm$ std over $K=5$ paired runs). \colorbox{green!35}{vision foundation models}; \colorbox{yellow!35}{radiomics baselines}. \emph{(ID)} rows: each cohort's test split. \emph{Cross-cohort}: averaged across all (train, eval) transfers between training cohorts. \emph{UCSD (ext)}: averaged across the four training-cohort sources.}\label{tab:calibration_ece}
    \resizebox{\textwidth}{!}{%
    \begin{tabular}{lcccccc}
        \toprule
        \multirow{2}{*}{\bfseries Setting}
          & \multicolumn{6}{c}{\bfseries ECE $\downarrow$} \\
        \cmidrule(lr){2-7}
          & \cellcolor{green!35}\bfseries BrainIAC
          & \cellcolor{green!35}\bfseries MRI-CORE
          & \cellcolor{green!35}\bfseries BiomedCLIP
          & \cellcolor{green!35}\bfseries BrainDINO
          & \cellcolor{yellow!35}\bfseries LogReg
          & \cellcolor{yellow!35}\bfseries TabPFN \\
        \midrule
        UCSF (ID)     & $0.28 \pm 0.07$ & $0.24 \pm 0.04$ & $0.09 \pm 0.04$ & $0.22 \pm 0.09$ & $0.26 \pm 0.02$ & $\mathbf{0.04 \pm 0.01}$ \\
        UPENN (ID)    & $0.32 \pm 0.18$ & $0.22 \pm 0.08$ & $0.09 \pm 0.05$ & $0.10 \pm 0.04$ & $0.30 \pm 0.16$ & $\mathbf{0.03 \pm 0.01}$ \\
        EGD (ID)      & $0.15 \pm 0.07$ & $0.20 \pm 0.05$ & $0.11 \pm 0.03$ & $0.10 \pm 0.03$ & $0.15 \pm 0.04$ & $\mathbf{0.09 \pm 0.02}$ \\
        UTSW (ID)     & $0.22 \pm 0.04$ & $0.19 \pm 0.08$ & $0.13 \pm 0.03$ & $0.12 \pm 0.02$ & $0.17 \pm 0.06$ & $\mathbf{0.10 \pm 0.03}$ \\
        Cross-cohort  & $0.24 \pm 0.13$ & $0.25 \pm 0.19$ & $0.14 \pm 0.07$ & $0.15 \pm 0.08$ & $0.18 \pm 0.09$ & $\mathbf{0.09 \pm 0.07}$ \\
        UCSD (ext)    & $0.30 \pm 0.12$ & $0.31 \pm 0.16$ & $0.13 \pm 0.07$ & $0.26 \pm 0.12$ & $0.18 \pm 0.09$ & $\mathbf{0.07 \pm 0.01}$ \\
        \bottomrule
    \end{tabular}%
    }
\end{table}

\section{Discussion}
Our central finding is that a tabular foundation model on handcrafted radiomics features matches or exceeds four frozen image foundation models on within-cohort IDH prediction across four glioma cohorts. This aligns with \textit{Nakase et al.} \cite{nakase_integration_2025}, extending their 158-patient observation to cohorts in the 500-patient range and to a recent tabular foundation model (TabPFN \cite{hollmann2023tabpfn}) rather than a classical classifier. Two of the three brain/MRI-specific encoders we tested (BrainIAC, MRI-CORE) underperform handcrafted radiomics on every cohort, while BrainDINO closes most of the gap to BiomedCLIP. This suggests that the SSL recipe (DINOv3 + iBOT on 6.6M brain MRI slices) matters at least as much as the imaging domain; domain-specific pretraining is not automatically a strong downstream starting point \cite{yuan_rethinking_2025}, but a properly designed self-supervised pretraining at scale on the target domain can be competitive with broader image–text pretraining. The picture changes under distribution shift. BiomedCLIP achieves the highest mean external AUROC on UCSD-PTGBM, while LogReg and TabPFN exhibit the largest drops from their source diagonal, plausibly because handcrafted descriptors are more sensitive than learned visual features to post-resection appearance changes. The within-cohort ranking is therefore a poor predictor of robustness to clinically relevant distribution shifts: the best in-distribution baseline is not the most robust baseline. Calibration sharpens this picture (Table~\ref{tab:calibration_ece}): TabPFN is the best-calibrated baseline in every setting, including UCSD-PTGBM (ECE $0.073$), so the BiomedCLIP AUROC advantage under severe shift does not translate into trustworthy probabilities. Since neither family is easily interpretable in a clinical sense (radiomics descriptors summarize the whole tumor and foundation-model embeddings are opaque by construction), calibration alongside ranking is a primary signal of deployment readiness: the right model choice depends on whether the downstream task needs accurate rankings (e.g.\ triage) or trustworthy probabilities (e.g.\ risk-threshold decisions). Several caveats temper these conclusions. All imaging foundation models are used through a frozen linear probe. Tumor segmentation source differs across cohorts (manual vs.\ automated), which may contribute to the cross-cohort transfer pattern. Finally, UCSD-PTGBM is restricted to the earliest available time point per patient, understating the shift relative to mid- or late-treatment imaging.

\section{Conclusion}
We show that, across four glioma cohorts and one external post-treatment cohort, radiomics features combined with a tabular foundation model yield the best in-distribution IDH prediction and competitive cross-cohort transfer. We further show that among image foundation models, BiomedCLIP and BrainDINO were the most competitive: BiomedCLIP's broad image–text pretraining gave it an edge under heavy distribution shift, while BrainDINO's brain-specific DINOv3 pretraining matched it within-cohort. The current MRI-specific encoders (BrainIAC, MRI-CORE) consistently underperformed, indicating that domain-specific pretraining is not categorically inferior, but only proves competitive when paired with a strong SSL recipe at scale. Across all settings, TabPFN remains the best-calibrated baseline, including under heavy post-treatment shift; ranking and calibration metrics therefore identify different winners depending on whether the downstream task needs accurate rankings or trustworthy probabilities. Several extensions follow naturally: richer probe architectures (e.g., cross-attention over FLAIR/T1 tokens) could close the within-cohort gap to TabPFN while preserving frozen-encoder robustness, whereas end-to-end fine-tuning risks eroding the pretrained features that underpin OOD performance \cite{kumar2022finetuningdistortpretrainedfeatures}. Joint use of imaging and clinical variables (age, WHO grade), larger post-treatment cohorts, and other molecular markers (1p/19q, MGMT, TERT) are natural next steps for testing whether our findings are IDH-specific or general across glioma genomics.

%
% ---- Bibliography ----
%
% BibTeX users should specify bibliography style 'splncs04'.
% References will then be sorted and formatted in the correct style.
%
\bibliographystyle{splncs04}
\bibliography{ref_v2}

@article{louis_2021_2021,
	title = {The 2021 {WHO} {Classification} of {Tumors} of the {Central} {Nervous} {System}: a summary},
	volume = {23},
	doi = {10.1093/neuonc/noab106},
	number = {8},
	journal = {Neuro-Oncology},
	author = {Louis, David N and others},
	month = jun,
	year = {2021},
	pages = {1231--1251},
}

@article{he_radiogenomics_2024,
	title = {Radiogenomics: bridging the gap between imaging and genomics for precision oncology},
	volume = {5},
	doi = {10.1002/mco2.722},
	number = {9},
	journal = {MedComm},
	author = {He, Wenle and Huang, Wenhui and Zhang, Lu and Wu, Xuewei and Zhang, Shuixing and Zhang, Bin},
	year = {2024},
	pages = {e722},
}

@article{an_radiomics_2021,
	title = {Radiomics machine learning study with a small sample size: {Single} random training-test set split may lead to unreliable results},
	volume = {16},
	doi = {10.1371/journal.pone.0256152},
	number = {8},
	journal = {PLOS ONE},
	author = {An, Chansik and Park, Yae Won and Ahn, Sung Soo and Han, Kyunghwa and Kim, Hwiyoung and Lee, Seung-Koo},
	month = aug,
	year = {2021},
	pages = {e0256152},
}

@article{sottoriva_intratumor_2013,
	title = {Intratumor heterogeneity in human glioblastoma reflects cancer evolutionary dynamics},
	volume = {110},
	doi = {10.1073/pnas.1219747110},
	number = {10},
	journal = {Proceedings of the National Academy of Sciences of the United States of America},
	author = {Sottoriva, Andrea and others},
	month = mar,
	year = {2013},
	pages = {4009--4014},
}

@article{farahani_diagnostic_2026,
	title = {Diagnostic performance of deep learning for predicting glioma isocitrate dehydrogenase and 1p/19q co-deletion in {MRI}: a systematic review and meta-analysis},
	volume = {36},
	doi = {10.1007/s00330-025-11898-2},
	number = {2},
	journal = {European Radiology},
	author = {Farahani, Somayeh and Hejazi, Marjaneh and Tabassum, Mehnaz and Di Ieva, Antonio and Mahdavifar, Neda and Liu, Sidong},
	month = feb,
	year = {2026},
	pages = {1562--1591},
}

@misc{kumar2022finetuningdistortpretrainedfeatures,
      title={Fine-Tuning can Distort Pretrained Features and Underperform Out-of-Distribution},
      author={Ananya Kumar and Aditi Raghunathan and Robbie Jones and Tengyu Ma and Percy Liang},
      year={2022},
      eprint={2202.10054},
      archivePrefix={arXiv},
      primaryClass={cs.LG},
      url={https://arxiv.org/abs/2202.10054},
}

@article{yuan_rethinking_2025,
	title = {Rethinking Domain-Specific Pretraining by Supervised or Self-Supervised Learning for Chest Radiograph Classification: A Comparative Study Against ImageNet Counterparts in Cold-Start Active Learning},
	volume = {4},
	doi = {10.1002/hcs2.70009},
	number = {2},
	journal = {Health Care Science},
	author = {Yuan, Han and Zhu, Mingcheng and Yang, Rui and Liu, Han and Li, Irene and Hong, Chuan},
	month = apr,
	year = {2025},
	pages = {110--143},
}

@misc{oquab2024dinov2learningrobustvisual,
      title={DINOv2: Learning Robust Visual Features without Supervision},
      author={Maxime Oquab and others},
      year={2024},
      eprint={2304.07193},
      archivePrefix={arXiv},
      primaryClass={cs.CV},
      url={https://arxiv.org/abs/2304.07193},
}

@misc{chen2020simpleframeworkcontrastivelearning,
      title={A Simple Framework for Contrastive Learning of Visual Representations},
      author={Ting Chen and Simon Kornblith and Mohammad Norouzi and Geoffrey Hinton},
      year={2020},
      eprint={2002.05709},
      archivePrefix={arXiv},
      primaryClass={cs.LG},
      url={https://arxiv.org/abs/2002.05709},
}

@article{isensee_automated_2019,
	title = {Automated brain extraction of multisequence {MRI} using artificial neural networks},
	volume = {40},
	doi = {10.1002/hbm.24750},
	number = {17},
	journal = {Human Brain Mapping},
	author = {Isensee, Fabian and others},
	month = aug,
	year = {2019},
	pages = {4952--4964},
}

@article{yan_idh1_2009,
	title = {{IDH1} and {IDH2} {Mutations} in {Gliomas}},
	volume = {360},
	doi = {10.1056/NEJMoa0808710},
	number = {8},
	journal = {The New England journal of medicine},
	author = {Yan, Hai and others},
	month = feb,
	year = {2009},
	pages = {765--773},
}

@article{lu_noninvasive_2023,
	title = {Noninvasive prediction of {IDH} mutation status in gliomas using preoperative multiparametric {MRI} radiomics nomogram: {A} mutlicenter study},
	volume = {104},
	doi = {10.1016/j.mri.2023.09.001},
	journal = {Magnetic Resonance Imaging},
	author = {Lu, Jun and Xu, Wenjuan and Chen, Xiaocao and Wang, Tan and Li, Hailiang},
	month = dec,
	year = {2023},
	pages = {72--79},
}

@article{choi_fully_2020,
	title = {Fully automated hybrid approach to predict the {IDH} mutation status of gliomas via deep learning and radiomics},
	volume = {23},
	doi = {10.1093/neuonc/noaa177},
	number = {2},
	journal = {Neuro-Oncology},
	author = {Choi, Yoon Seong and others},
	month = jul,
	year = {2020},
	pages = {304--313},
}

@article{calabrese_combining_2022,
	title = {Combining radiomics and deep convolutional neural network features from preoperative {MRI} for predicting clinically relevant genetic biomarkers in glioblastoma},
	volume = {4},
	doi = {10.1093/noajnl/vdac060},
	number = {1},
	journal = {Neuro-oncology Advances},
	author = {Calabrese, Evan and Rudie, Jeffrey D and Rauschecker, Andreas M and Villanueva-Meyer, Javier E and Clarke, Jennifer L and Solomon, David A and Cha, Soonmee},
	month = apr,
	year = {2022},
	pages = {vdac060},
}

@misc{zhang_biomedclip_2025,
	title = {{BiomedCLIP}: a multimodal biomedical foundation model pretrained from fifteen million scientific image-text pairs},
	doi = {10.48550/arXiv.2303.00915},
	publisher = {arXiv},
	author = {Zhang, Sheng and others},
	month = jan,
	year = {2025},
}

@misc{dong_mri-core_2025,
	title = {{MRI}-{CORE}: {A} {Foundation} {Model} for {Magnetic} {Resonance} {Imaging}},
	doi = {10.48550/arXiv.2506.12186},
	publisher = {arXiv},
	author = {Dong, Haoyu and Chen, Yuwen and Gu, Hanxue and Konz, Nicholas and Chen, Yaqian and Li, Qihang and Mazurowski, Maciej A.},
	month = jul,
	year = {2025},
}

@article{tak_generalizable_2026,
	title = {A generalizable foundation model for analysis of human brain {MRI}},
	volume = {29},
	doi = {10.1038/s41593-026-02202-6},
	number = {4},
	journal = {Nature Neuroscience},
	author = {Tak, Divyanshu and others},
	month = apr,
	year = {2026},
	pages = {945--956},
}

@article{bakas_university_2022,
	title = {The {University} of {Pennsylvania} glioblastoma ({UPenn}-{GBM}) cohort: advanced {MRI}, clinical, genomics, \& radiomics},
	volume = {9},
	doi = {10.1038/s41597-022-01560-7},
	number = {1},
	journal = {Scientific Data},
	author = {Bakas, Spyridon and others},
	month = jul,
	year = {2022},
	pages = {453},
}

@article{calabrese_university_2022,
	title = {The {University} of {California} {San} {Francisco} {Preoperative} {Diffuse} {Glioma} {MRI} ({UCSF}-{PDGM}) {Dataset}},
	volume = {4},
	doi = {10.1148/ryai.220058},
	number = {6},
	journal = {Radiology: Artificial Intelligence},
	author = {Calabrese, Evan and others},
	month = nov,
	year = {2022},
	pages = {e220058},
}

@article{van_der_voort_erasmus_2021,
	title = {The {Erasmus} {Glioma} {Database} ({EGD}): {Structural} {MRI} scans, {WHO} 2016 subtypes, and segmentations of 774 patients with glioma},
	volume = {37},
	doi = {10.1016/j.dib.2021.107191},
	journal = {Data in Brief},
	author = {van der Voort, Sebastian R. and others},
	month = jun,
	year = {2021},
	pages = {107191},
}

@article{nakase_integration_2025,
	title = {Integration of {MRI} radiomics and germline genetics to predict the {IDH} mutation status of gliomas},
	volume = {9},
	doi = {10.1038/s41698-025-00980-z},
	number = {1},
	journal = {npj Precision Oncology},
	author = {Nakase, Taishi and others},
	month = jun,
	year = {2025},
	pages = {187},
}

@article{ziv_importance_2016,
	title = {The {Importance} of {Biopsy} in the {Era} of {Molecular} {Medicine}},
	volume = {22},
	doi = {10.1097/PPO.0000000000000228},
	number = {6},
	journal = {Cancer journal (Sudbury, Mass.)},
	author = {Ziv, Etay and Durack, Jeremy C. and Solomon, Stephen B.},
	year = {2016},
	pages = {418--422},
}

@article{van_der_heide_mri_2019,
	title = {{MRI} basics for radiation oncologists},
	volume = {18},
	doi = {10.1016/j.ctro.2019.04.008},
	journal = {Clinical and Translational Radiation Oncology},
	author = {van der Heide, Uulke A. and Frantzen-Steneker, Marloes and Astreinidou, Eleftheria and Nowee, Marlies E. and van Houdt, Petra J.},
	month = apr,
	year = {2019},
	pages = {74--79},
}

@misc{bommasani_opportunities_2022,
	title = {On the {Opportunities} and {Risks} of {Foundation} {Models}},
	doi = {10.48550/arXiv.2108.07258},
	publisher = {arXiv},
	author = {Bommasani, Rishi and others},
	month = jul,
	year = {2022},
}

@inproceedings{qin_l1-2_2019,
	title = {L1-2 {Regularized} {Logistic} {Regression}},
	doi = {10.1109/IEEECONF44664.2019.9048830},
	booktitle = {2019 53rd {Asilomar} {Conference} on {Signals}, {Systems}, and {Computers}},
	author = {Qin, Jing and Lou, Yifei},
	month = nov,
	year = {2019},
	note = {ISSN: 2576-2303},
	pages = {779--783},
}

@misc{radford_learning_2021,
	title = {Learning {Transferable} {Visual} {Models} {From} {Natural} {Language} {Supervision}},
	doi = {10.48550/arXiv.2103.00020},
	publisher = {arXiv},
	author = {Radford, Alec and others},
	month = feb,
	year = {2021},
}

@misc{reddy_university_2026,
	title = {The {University} of {Texas} {Southwestern} {Glioma} {MRI} dataset with molecular marker characterization and segmentations ({UTSW}-{Glioma})},
	doi = {10.7937/DFAE-1B86},
	publisher = {The Cancer Imaging Archive},
	author = {Reddy, D. and others},
	year = {2026},
	note = {Version 1},
}

@misc{gagnon_university_2025,
	title = {The {University} of {California} {San} {Diego} annotated post-treatment high-grade glioma multimodal {MRI} dataset ({UCSD}-{PTGBM})},
	doi = {10.7937/FWV2-DT74},
	publisher = {The Cancer Imaging Archive},
	author = {Gagnon, L. and others},
	year = {2025},
	note = {Version 3},
}

@inproceedings{hollmann2023tabpfn,
  title={TabPFN: A transformer that solves small tabular classification problems in a second},
  author={Hollmann, Noah and M{\"u}ller, Samuel and Eggensperger, Katharina and Hutter, Frank},
  booktitle={International Conference on Learning Representations 2023},
  year={2023}
}

@misc{paszke2019pytorchimperativestylehighperformance,
      title={PyTorch: An Imperative Style, High-Performance Deep Learning Library},
      author={Paszke, Adam and others},
      year={2019},
      url={https://arxiv.org/abs/1912.01703},
}

@misc{wu2026braindinobrainmrifoundation,
      title={BrainDINO: A Brain MRI Foundation Model for Generalizable Clinical Representation Learning}, 
      author={Yizhou Wu and Shansong Wang and Yuheng Li and Mojtaba Safari and Mingzhe Hu and Chih-Wei Chang and Harini Veeraraghavan and Xiaofeng Yang},
      year={2026},
      eprint={2604.27277},
      archivePrefix={arXiv},
      primaryClass={cs.LG},
      url={https://arxiv.org/abs/2604.27277}, 
}

@misc{simeoni2025dinov3,
      title={DINOv3},
      author={Siméoni, Oriane and others},
      year={2025},
      url={https://arxiv.org/abs/2508.10104},
}

\end{document}